\documentclass[twocolumn,prl,amsmath,floatfix]{revtex4}
\usepackage{graphicx}
\usepackage{epsfig}

\begin{document}
\title{Force induced triple point for interacting polymers} 
\author{Sanjay Kumar and  Debaprasad Giri}
\affiliation{Department of Physics, Banaras Hindu University,
     Varanasi 221 005, India}
\email{SK: yashankit@yahoo.com; DG: dgiri99@yahoo.com}
\author{Somendra M. Bhattacharjee}
\affiliation{Institute of Physics, Bhubaneswar 751 005, India}
\email{somen@iopb.res.in}
\date{\today}
\begin{abstract}
  We show the existence of a force induced triple point in an
  interacting polymer problem that allows two zero-force thermal phase
  transitions.  The phase diagrams for three different models of
  mutually attracting but self avoiding polymers are presented.  One
  of these models has an intermediate phase and it shows a triple
  point but not the others.  A general phase diagram with
  multicritical points in an extended parameter space is also
  discussed.
\end{abstract}

\maketitle

\newcommand{\figone}{%
\begin{figure}[htbp]
{\includegraphics[width=2.in,clip]{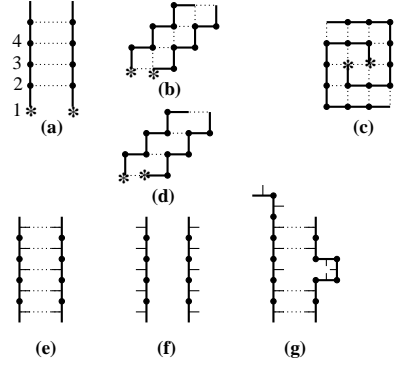}}
\caption{ Figures (a-d) represent the possible conformations of models A and B. 
  In model A any monomer of one strand can interact with any monomer
  of the other strand. (a) and (b) are two possible states with (c)
  representing a possible ground state.  For model B, the $i$th
  monomer of one strand interacts with the $i$th monomer of the other
  strand.  In this case (a) represents the ground state, (d)
  represents a partial bound state.  Note that (d) differs from (b) in
  the nature of interactions represented by the dotted lines.  In
  model B, (c) has no valid interaction and would represent an open
  state.  Figures (e,f,g) represent model C where bases are on the
  links of the strands with short stubs representing orientation.  (e)
  the completely zipped state, (f) a non pairing configuration, and
  (g) a partial bound configuration.  In all cases dotted lines show
  the attractive interactions.}
\label{fig-1}
\end{figure}%
}

\newcommand{\figtwo}{%
\begin{figure}
\includegraphics[width=1.5in,clip]{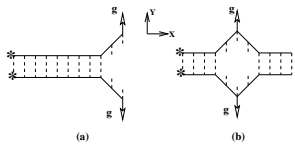}
\caption{Schematic figures for  a force (a)  applied
  at one end, and (b) applied at the middle. Fixed end points are
  indicated by stars.}
\label{fig-2}
\end{figure}%
}

\newcommand{\figthree}{%
\begin{figure}
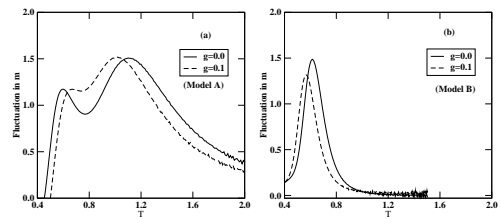

\includegraphics[width=1.25in,clip]{fig3a.eps}
\includegraphics[width=1.25in,clip]{fig3b.eps}
\caption{Variations in fluctuation of $m$ with temperature ($T$) for $g=0.0$ and $0.1$
  are shown for model A (a) and model B (b).  In model B and C we do not find two peaks.
  The first peak in (a)  corresponds to a transition where spiral
  conformation goes over to a zipped conformation. The second peak in
  (a)  and the peak in (b) 
  correspond to the unzipping transition. }
\label{fig-3}
\end{figure}%
}

\newcommand{\figfour}{%
\begin{figure}
\includegraphics[width=1.5in,clip]{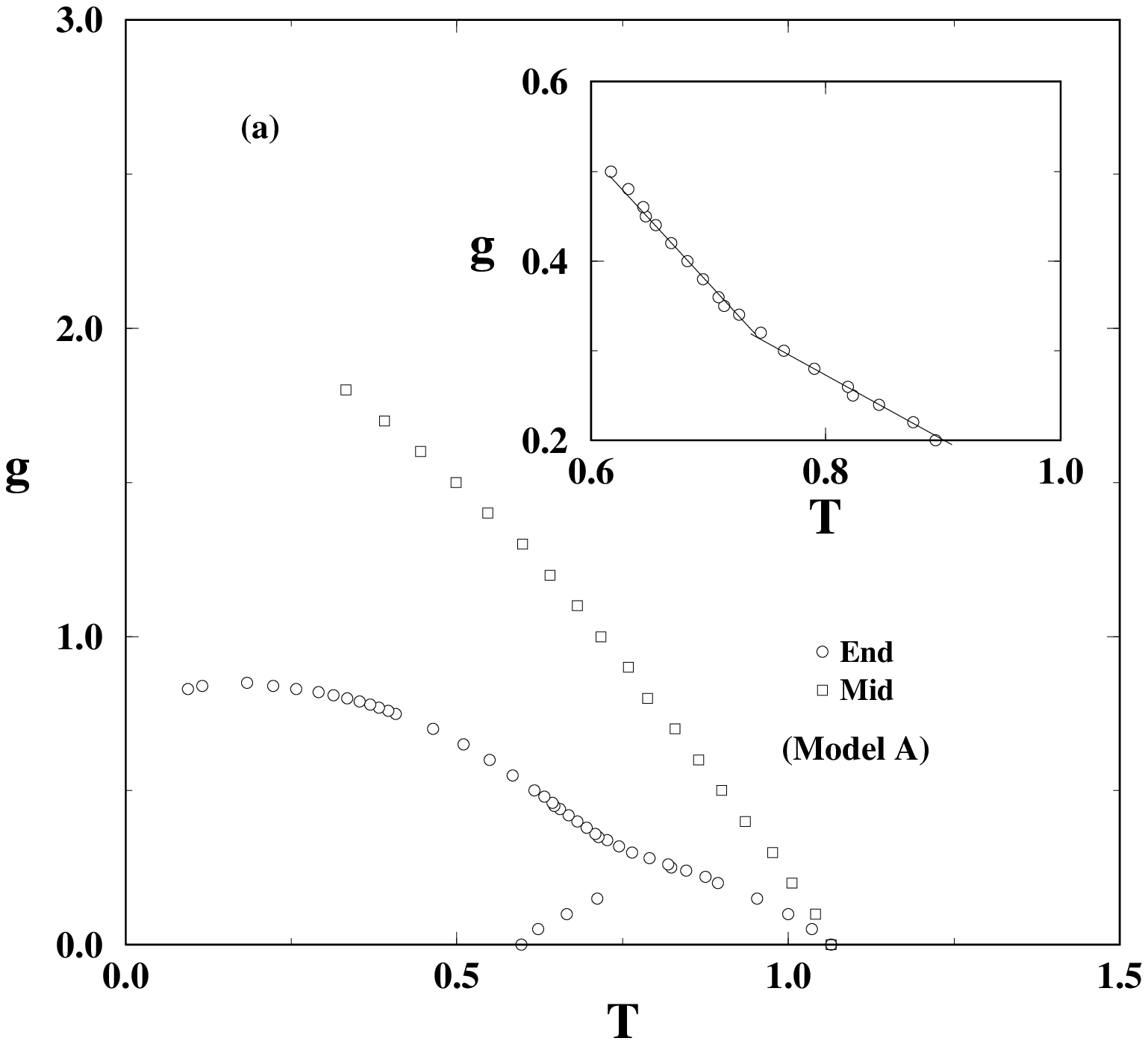}
\includegraphics[width=1.5in,clip]{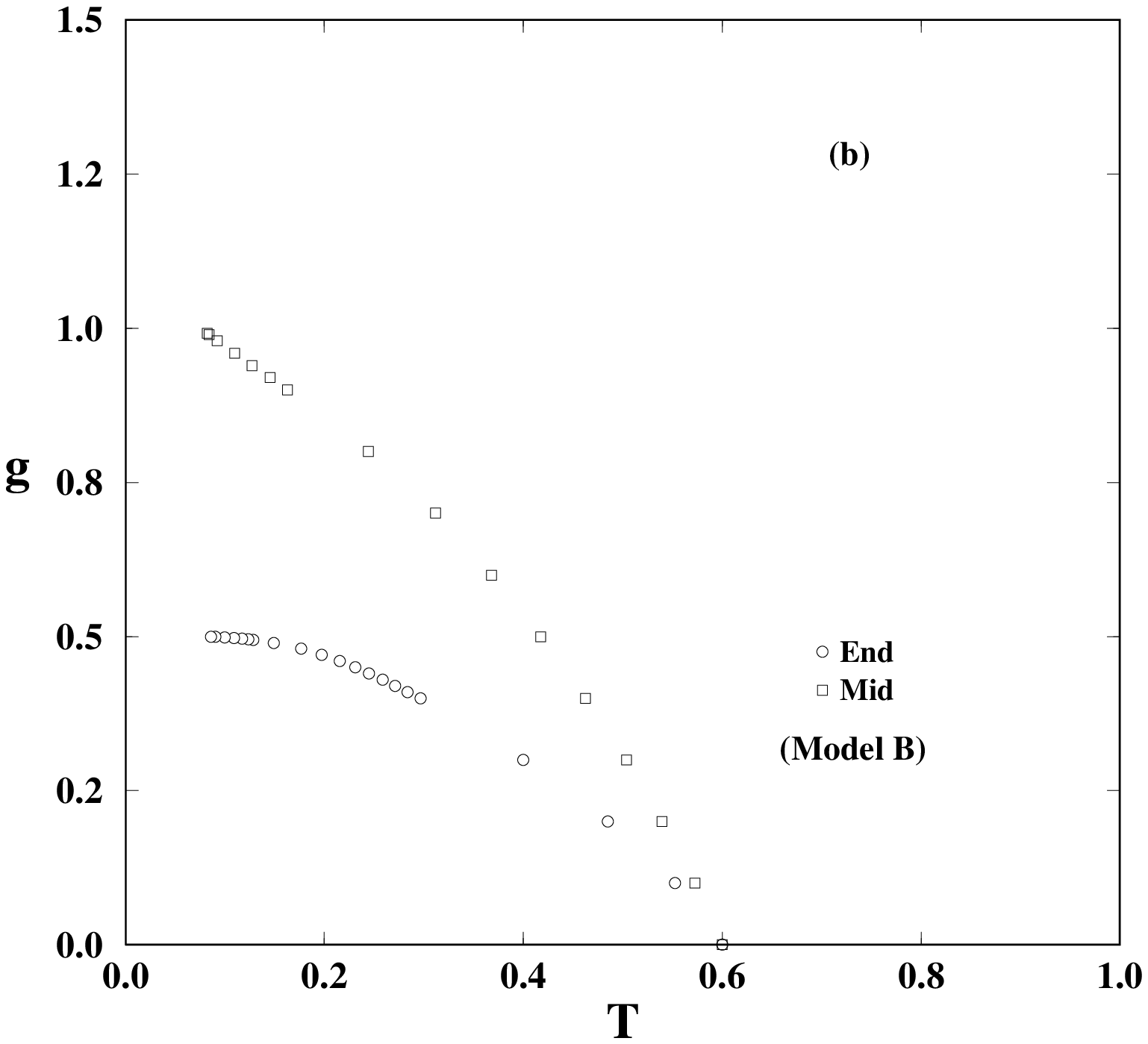}
\caption{Force-Temperature (g-T) phase diagrams for the end and the mid case. 
  (a) for model A, (b) model B and model C within errorbar. }
\label{fig-4}
\end{figure}%
}

\newcommand{\figfive}{%
\begin{figure}
\includegraphics[width=2.5in,clip]{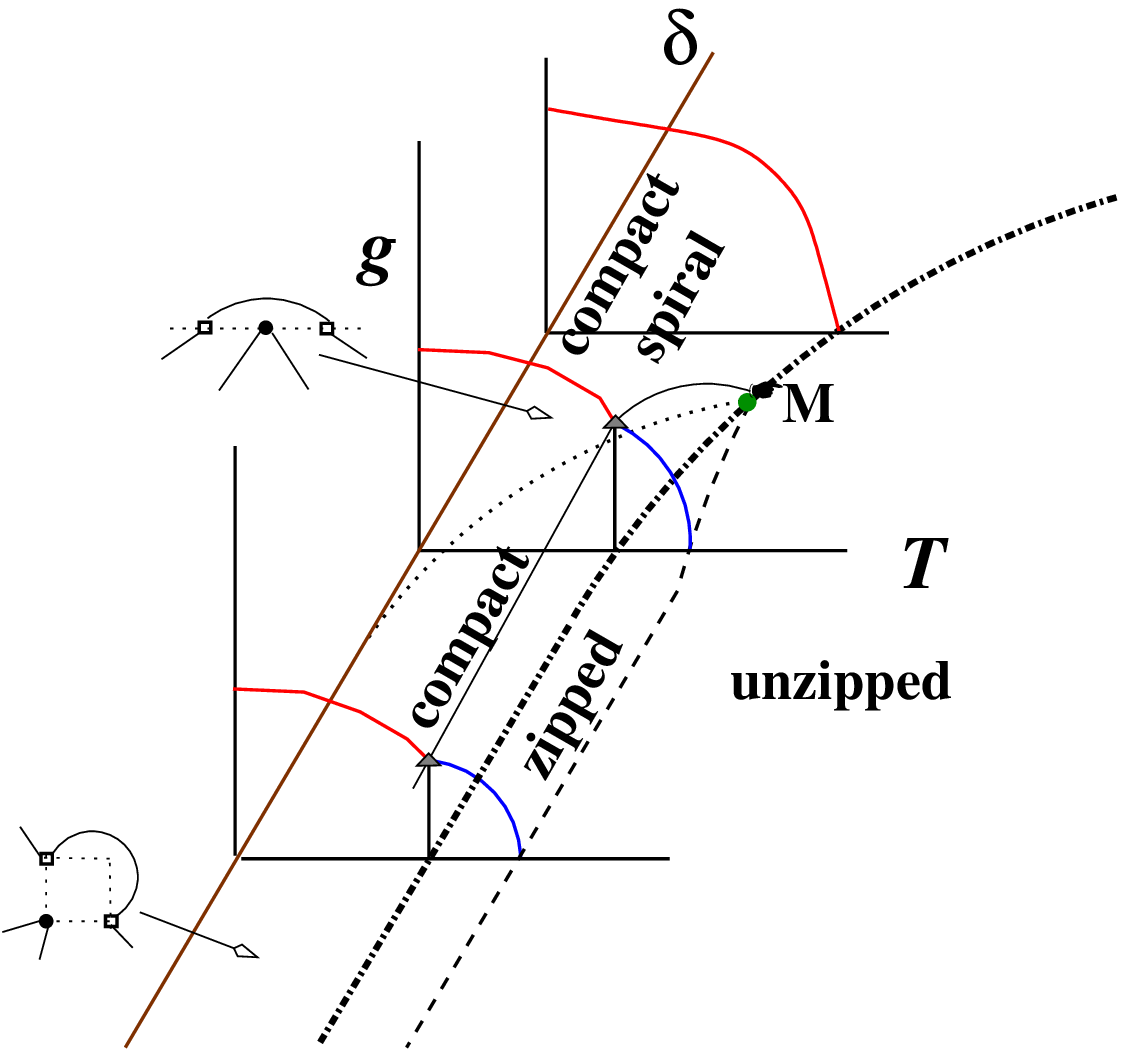}
\caption{Proposed phase diagram in $g-T-\delta$ space.  M is the
  multicritical point (at zero force) where the zipped phase vanishes
  as $\delta$ is made negative (attractive three-body interaction).
  The dotted line in the $g=0$ plane is a crossover line for a
  $\delta$-interaction dominated phase to go over a non-spiral compact
  phase dominated by interactions at corners. The triangular region in
  the $g-T$ plane for the zipped phase vanishes at M.  M is therefore
  the end point of the line of triple points (thin line through the
  triangles).}
\label{fig-5}
\end{figure}%
}
\newcommand{\figsix}{%
\begin{figure}
\includegraphics[width=2.5in,clip]{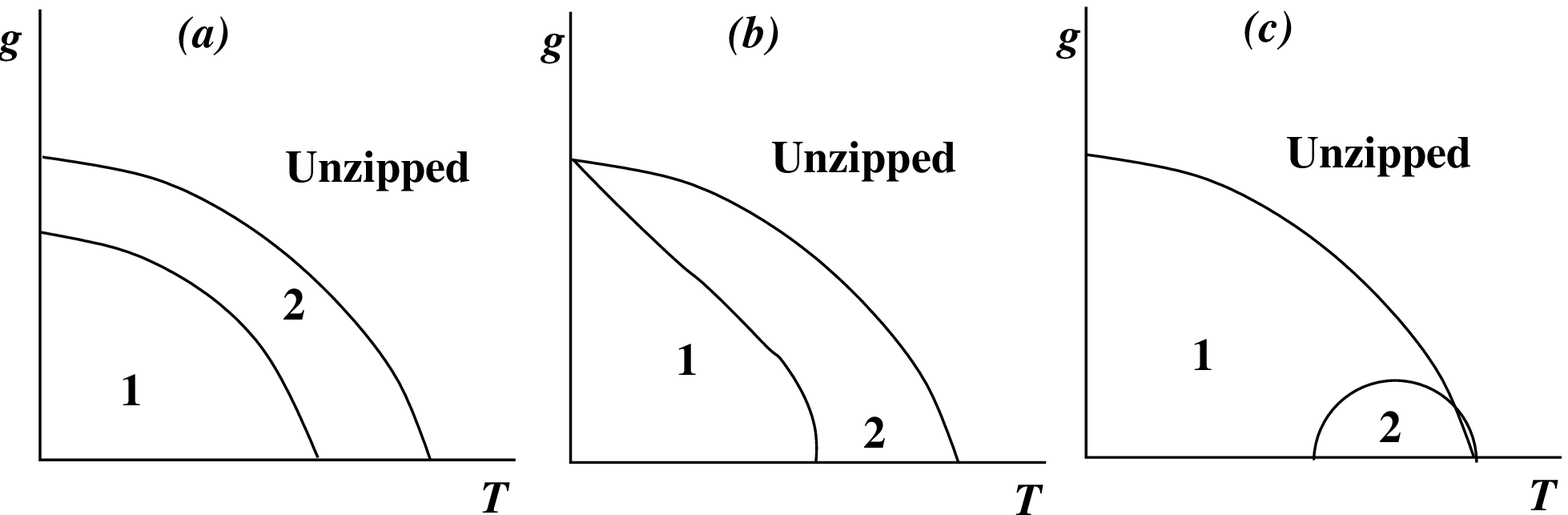}
\caption{Possible phase diagrams {\it without any triple point} in the $g-T$
  plane.  Phase 2 (a) may persist till $T=0$ under force, (b) may
  persist under force but disappears at $T=0$, or (c) exists only over
  a limited range of force and temperature.  In (a) the sequence of
  1-2-unzipped transitions will be observable in the groundstate
  itself, while in (c) a reentrant behaviour will be observable within
  phase 1 itself.  }
\label{fig-6}
\end{figure}%
}

Soon after the discovery of the double helical structure of double
stranded DNA (dsDNA), its melting by change of temperature or the pH
of the solvent was recognized\cite{doty}.  Only recently it has been
realized that there can be a force-induced unzipping
transition\cite{bhat99} of a DNA with a force applied solely at one
end.  In both the thermal and the forced cases, the phase transition
takes a double stranded form to two single strands.  The prediction of
the unzipping transition based on interacting Gaussian
chains\cite{bhat99} was immediately reconfirmed by a dynamical
approach\cite{seba}.  Results on the unzipping transition are now
available from extensive exact solutions of lattice
models\cite{trieste,maren2k2,kbs}, simple models of quenched-averaged
DNA \cite{lam}, Monte Carlo simulations of a three dimensional model
with self- and mutually-avoiding walks \cite{orland} etc., on the
theoretical front, and for real DNA from experiments\cite{dani}.  From
the theoretical results, it emerges that the qualitative features of
the unzipping transition are insensitive to the dimensionality ($d$)
of the models and are seen even in two dimensional models.  These
results include a re-entrance\cite{trieste,maren2k2,kbs} in the low
temperature region.

In the systems studied so far, there is only one zero force thermal
phase transition.  In such cases, with two intensive variables,
temperature $T$ and force $g$, there is an unzipping transition line,
$g=g_c(T)$, in the $g- T$ plane, demarcating the bound or zipped phase
from the unzipped phase.  This gives the unzipping phase boundary.  In
case there are more than one transition, the phase diagram will be
influenced by the intermediate phases.  Our aim is to determine such a
global phase diagram for the unzipping transition where the existence
of an intermediate phase leads to a triple point in the $g-T$ plane.

\figone

We introduce three models, A, B and C, which differ in the nature of
the mutual interaction of the two strands.  These models are general
enough to be defined in any $d$ though, for computational limitations,
we consider the two dimensional square lattice version only.  Let us
consider two linear polymer chains which are
mutually-attracting-self-avoiding \cite{exenu} in nature.  On a square
lattice, the polymers are not allowed to cross each other as shown in
Fig.  \ref{fig-1}.  Monomers are the sites occupied by the polymers
and the interactions are among the ``bases'' (defined next) so that an
interacting pair may also be called a base pair.  In models A and B,
the bases are the monomers while in C the bases are associated with
the links.  We allow an attractive interaction between monomers or
bases only if they are of opposite strands and are nearest neighbours
on the lattice.  The nearest neighbour interaction mimics the short
range nature of the hydrogen bonds.  In model A, any monomer of one
strand can interact with any monomer of the other strand. In model B,
monomer $i$ of one strand can interact only with the $i$th monomer of
the other strand.  This model is similar to the models of DNA studied
earlier\cite{pspb}.  In model C, we associate the bases with the bonds
of the strands. In one strand the bases point towards the right while
on the other they are on the left, as one traverses the chains
sequentially\cite{comm1}.  Pairing between complimentary strands takes
place only when the bases approach each other directly without the
strands coming in-between, as shown in Fig.  \ref{fig-1}).  Such a
model has been studied recently in the context of relative stabilities
of DNA hairpin structures\cite{gks}.  The figure also shows a
situation where pairing cannot take place and such configurations are
counted in the unbound states.  The interactions are shown by dotted
lines in Fig.  \ref{fig-1}).  The interaction energy in all the cases
is taken as $-\epsilon (\epsilon>0)$ and we shall choose $\epsilon=1$.
In all cases the monomers at one end (index=1) of each strand are
always kept fixed occupying nearest neighbour sites on the lattice.

For model A the ground state is a spiral of the type shown in Fig.
\ref{fig-1}c while for model B and C it is a zipped state as in Fig.
\ref{fig-1}a,e.  Because of the constraint of holding the monomers
with index 1 at nearest neighbour sites, model A is also equivalent to
a diblock copolymer model\cite{baiesi} which has two phase transitions
in zero force with increasing temperature.  The polymers go from a
compact spiral-like phase to a zipped phase which then melts at a
still higher temperature.  It is this intermediate phase that is of
interest to us.  In view of the simple ground state in models B and C,
no such intermediate phases are expected or known.

A force is applied at one end (Fig.\ref{fig-2}(a)) or at the middle
(Fig. \ref{fig-2}(b)) of the chains, in the $y$-direction.  The
contribution to energy by this force is $-g y$ where $y$ is the
absolute distance in the $y$-direction between the two strands at the
point of application of the force.  A recent study\cite{kbs} showed a
rich phase diagram when a force is applied somewhere in the interior.
Furthermore, such situations occur in many processes like
gene-expression where RNA is formed in bubbles or eye-type
configurations on DNA.  Motivated by these, we consider the case of a
force applied at the mid point.  A low temperature analysis shows that
the phase diagram is expected to be different from the end case.  We
compare the $g-T$ phase diagrams of model A vs models B and C.

\figtwo

The thermodynamic properties associated with the unzipping transition
are obtained from the partition function which can be written as a sum
over all possible configurations
\begin{eqnarray}
Z_N(\omega,u)    & = & \sum_{m,y} C(m,y)  \omega^m u^y.
\end{eqnarray} 
Here $N$ is the chain length of each of the two strands, 
$\omega=\exp(1/T)$ is the Boltzmann weight associated with each base
pair (taking the Boltzmann constant $k_{\rm B}=1$) and $m$ is the
total number of intact base pairs in the chain. Finally $u$ is the
Boltzmann weight, $\exp(g/T)$ associated with force.  $C(m,y)$ is the
number of distinct configurations having $m$ base pairs whose end (or
mid) points are at a distance $y$ apart. We have obtained $C(m,y)$ for
$N\le 16$ monomers in two dimension ($d=2$) and analyzed the partition
function through series analysis.  We prefer this technique because in
this case the scaling corrections are correctly taken into account by
suitable extrapolation technique.  To achieve the same accuracy by the
Monte-Carlo method, a chain of about two orders of magnitude larger
than in the exact enumeration method has to be considered
\cite{pgrass}.

The reduced free energy per base pair is found from the relation
\begin{equation}
G(\omega,u)= \lim_{N \rightarrow \infty} \frac{1}{N} \log Z(\omega,u).
\end{equation}
The limit $N \rightarrow \infty$ is found by using the ratio method
\cite{ratio} for extrapolation. The transition point for zero force
(i.e.  thermal melting) can be obtained from the plot of $G(\omega,u)$
versus $\omega$ or from the peak value of ${\partial^2 G}/{\partial
  (\ln\omega)^2}$.  For self-avoiding walk at $u=1$, we find $T_c=
1.1\pm 0.1, 0.61 \pm 0.08, 0.59\pm 0.08$ for models A, B and C
respectively.  With a force ($u\neq 1$), the phase boundary is
obtained from the fluctuation in $m$.  Fig. \ref{fig-3} shows the
variation of fluctuation of $m$ with temperature for model A and B.

\figthree

The phase diagrams are shown in Fig. \ref{fig-4}.

The phase diagram for Model A has several distinctive features.  We
see two peaks in the temperature dependence of the fluctuation in $m$
for small $g$.  At low $T$ there is a spiral structure with vanishing
entropy but as $T$ increases ($T>0.5$ ) the chains acquire a zipped
state, eventually unzipping at higher temperatures.  The low
temperature peak is the transition where the spiral state goes over to
the zipped state while the second peak is the unzipping transition.
Although model C is a more realistic model for representation of
dsDNA, we find the phase diagram to be almost the same as that of
model B.  This may be taken as an {\i a posteriori} justification of
using model B for DNA.  (However, in case of hetrogeneous chains, model
C may give intermediate states by forming the hairpin kind of structure
as shown in Fig1 e. Such structures are not possible in model B.)

Our results are based on the $N=16$ (32 monomers) enumerations.  For
this length there is significant surface contribution.  The ground
state energy is $E_0 = -(2N-{\mathrm O}({\sqrt{N}}))$ while
the ground state energy is $E_0=-N\epsilon$ for both models B and C
for a DNA with $N$ monomers or bases each.  The ${\mathrm
  O}({\sqrt{N}})$ correction for model A comes from the monomers on
the boundary.  If we ignore the surface contribution (valid for large
$N$) the spiral to zipped state transition temperature $T_{c1}$ may be
estimated from a simple energy balance argument.  The spiral state has
an energy $E_0=-2N$ with negligible entropy while the zipped
state has the free energy $=-N - N T\ln \mu_b$, where $\ln
\mu_b$ is the entropy per base pair of the zipped phase. Equating
these two we get
\begin{equation}
  \label{eq:3}
  T_{c1}={1}/{\ln \mu_b} -  O(N^{-1/2} ),
\end{equation}
where the surface correction has also been shown.  If we use
$\mu_b=2.682$, the connectivity constant for the square lattice
self-avoiding walk\cite{exenu}, we find $T_{c1}\approx 1.04$  which is
close to the value known from other estimates\cite{baiesi}.  Since 
the surface correction  lowers the estimate of $T_{c1}$, our value of
$T_{c1}=0.5$ is  consistent with the form of Eq. \ref{eq:3}.

\figfour

The surface contribution ($O({\sqrt{N}})$ in the spiral state for
small chains is not negligible, and it helps in the stabilization of
of the spiral structure over the zipped phase for small $g$ because of
the extra energy $\sim g {\sqrt{N}}$.  As a result, the phase boundary
obtained has a finite slope.  For large $N$, the force term should not
affect the spiral to zipped transition since the force affects the
relative separation of the strands.  Therefore one would expect a
phase boundary parallel to the $g$-axis {\it meeting the unzipping
  phase boundary at a triple point}.  Other possible phase diagrams
are shown in Fig. \ref{fig-6}. From thermodynamic stability
analysis\cite{wheeler}, it is known that the angle between two
coexistence lines at a triple point in a phase diagram has to be less
than $\pi$.  Therefore, a discontinuity in the slope in the unzipping
phase boundary in the $g-T$ plane is expected at the potential triple
point.  The inset in Fig.  \ref{fig-4}a shows the meeting of the two
unzipping boundaries.  Our main result is the prediction of
the coexistence of all the three phases in the $g-T$ phase diagram for
model A - a phenomenon not possible without force.  

At $T=0$, the critical force can be found from a matching of the
ground state energy with the energy of the completely unzipped state.
If the force stretches the strands completely, the unzipped state has
the energy $=-2N g$ taking the bond length to be unity.  Comparing
this with the bound state energy $E_0$, we see $g_c(T=0)= E_0/2N$.
For large $N$, $E_0=2N$ and so we get $g_c(T=0)=1$, while it is $0.5$
for models B and C.  For finite $N$, if the surface correction is
taken into account, then $g_c=1-(1/{\sqrt{2N}}) \approx 0.8232$ for
$N=16$.  The value shown in Fig. \ref{fig-4} is very close to this
estimate rather than the large $N$ value.  Any extrapolation therefore
would have to take into account this surface correction properly.  In
absence of any large $N$ data we do not attempt such extrapolations
here for model A.   

The low temperature phase boundary for the mid case can be obtained by
an extension of the $T=0$ argument given above.  Following Ref.
\cite{kbs}, let us consider the situation with the force applied at a
position $sN$, $(0<s\le 1)$, from the fixed end.  The unzipped state
has the energy $=-2sN g$ taking the bond length to be unity.
Comparing this with the bound state energy we see $g_c(T=0)= a
/2s$ where $a=2$ for model A but $a=1$ for models B and C.  We
see a factor of $2$ difference in the end vs mid case.

\figsix

Close to $T=0$, the free energy  of a bubble of $2m$ base pairs with
respect to the completely bound state is 
\begin{equation}
  \label{eq:1}
  \Delta F = 2m(a  + T \ln \mu_b) - g m,
\end{equation}
where $\ln \mu_b$ is the entropy per base pair of the bound state which
is lost on unzipping $m$ pairs.  For the mid case  a large $m$ is
favourable ($m\leq N/2$) if $g>g_c(T)\equiv 2(a+ T \ln \mu_b)$.
In case $\ln \mu_b > 0$, we see a positive slope and hence a
re-entrance, in the low temperature region. In all the three models
studied here, $\ln \mu_b=0$, and therefore no re-entrance.

\figfive 

The results for model A can be extended to a case with a three body
interaction that stabilizes the spiral phase\cite{baiesi}.  Let there
be an interaction $\delta$ that favours a configuration with a monomer
of one strand sandwiched between the monomers of the other strand on
the two nearest neighbors on the same axis. Let us call such contact
as $\delta$-contact.  For $\delta<0$ (i.e. attractive three body
interaction) the compact spiral phase is the ground state as at
$\delta=0$.  For large negative $\delta$, there is only one
denaturation transition.  Therefore a multicritical point occurs in
the $g=0 (\delta-T)$ plane.  This is indicated by M in Fig.
\ref{fig-5}.  For repulsion, $\delta>0$, one may have a compact ground
state without any $\delta$-contact.  In this case the ground state
need not be a spiral and will have a mixture of double contacts (e.g. at
corners) and single contacts.  Let these occur in proportion $p:(1-p)$
($p<1$).  The ground state energy is then $-N(1+p)$.  A
meanfield estimate of equal proportion gives $E=-(3/2)N$.
This energy needs to be compared with the energy of the spiral state,
namely $-2N+2N\delta$ for $N\rightarrow\infty$.  At $T=0$, a
change in the ground state structure occurs at $\delta= 1-p\approx
0.5$.  In the $\delta-T$ phase diagram one will have a crossover line
(dotted line in Fig.  \ref{fig-5}) which should end in the $\delta-T$
plane at the multicritical point which is the confluence of the two
zero-force thermal transitions.  For large negative $\delta$, the
compact phase goes over directly to the unzipped phase and therefore
no triple point can exist on the unzipping boundary.  The $g-T-\delta$
phase diagram is schematically shown in FIg. \ref{fig-5}.

To summarize, we have shown in this paper the existence of a
force induced triple point in a problem of interacting polymers where
the interactions allow an entropy-stabilized phase before melting.
This opens up the possibility of much richer phase transitions and
other multicritical behaviours in polymeric systems  when subjected to
force.    We hope that single molecule
experiments would be able to explore experimentally this new area in polymers.

SK and DG would like to thank Y. Singh for many helpful
discussions on the subject  and acknowledge  financial assistance from
INSA (New Delhi) and DST (New Delhi).


\begin{thebibliography}{99}
\bibitem{doty} P. Doty, J. Marmur, J. Eigner and C. Schildkraut.
Proc. Nat. Acad. Sci. (U. S. A.) 46, 453, 461 (1960)
\bibitem{bhat99} S. M. Bhattacharjee, J. Phys. A {\bf 33} L423 (2000) (cond-mat/9912297).
\bibitem{seba} K. Sebastian, Phys. Rev. E 62, 1128 (2000).
\bibitem{trieste} D. Marenduzzo, A. Trovato, and A. Maritan,
Phys. Rev. E 64, 031901 (2001).  
\bibitem{maren2k2} D. Marenduzzo, S. M. Bhattacharjee, A. Maritan,
E. Orlandini  and F. Seno,  Phys. Rev. Lett. 88, 028102 (2002)  
\bibitem{kbs} R. Kapri, S. M. Bhattacharjee and F. Seno,  cond-mat/0403752 
\bibitem{lam} Puri-Man Lam {\it et~al.}, Biopolymers {\bf 73}, 293 (2004);
D. K. Lubensky and D. R. Nelson, Phys. Rev. Lett. 85, 1572 (2000);
Jeff  Z.Y. Chen, Phys. Rev. {\bf E 66}, 031912 (2002) . 
\bibitem{orland} E. Orlandini et al,  J. Phys. {\bf A 34}, L751 (2001). 
\bibitem{dani} C. Danilowicz et al,  cond-mat/0310633 
\bibitem{exenu} T. Ishinabe, J. Chem. Phys {\bf 76}, 5589 (1982); 
C. Vanderzande, {\it Lattice models of polymers} (Cambridge University 
Press, UK, 1998)
\bibitem{pspb} D. Poland and H. Scheraga, J. Chem. Phys. 45, 1464
  (1966).
\bibitem{comm1} Note that real DNA
  strands have opposite polarity $3^{'}-5^{'} $ for one but
  $5^{'}-3^{'}$ for the other.  Such differences are of no concern to
  the models of DNA considered so far and to be discussed here.
\bibitem{gks} D. Giri, S. Kumar and Y. Singh, to be published
\bibitem{baiesi}  E. Orlandini, F. Seno, and A.L. Stella,
  Phys. Rev. Lett. 84, 294 (2000).; M. Baiesi et al, Phys. Rev. E {\bf 63}, 041801 (2001).
\bibitem{pgrass} P. Grassberger and R. Hegger  {Phys. Rev.} {\bf E51} 
2674 (1995); P. Grassberger and R. Hegger {J. Physique I.} France {\bf 5} 
597 (1995)
\bibitem{ratio}A. J. Guttmann {\it Phase Transition and Critical Phenomena},
edited by C. Domb and J. L. Lebowitz  (Academic, New York) {\bf Vol 13} 
(1989);  R. Rajesh, D. Dhar, D. Giri, S. Kumar and Y. Singh, Phy.Rev. E
056124 (2002)
\bibitem{wheeler} J. Wheeler, J. Chem. Phys. {\bf 61}, 4474 (1974).
\end{thebibliography}
\end{document}